\documentclass[prd,amssymb,amsmath,amsfonts,superscriptaddress,nofootinbib,reprint,showpacs, longbibliography]{revtex4-1}

\usepackage{graphicx}
\usepackage{lmodern}
\usepackage{amsmath,amssymb}
\usepackage{mathrsfs}
\usepackage{amsfonts}
\usepackage[utf8]{inputenc}
\usepackage{url}
\usepackage[colorlinks]{hyperref}
\usepackage[dvipsnames]{xcolor}
\usepackage{multirow}
\usepackage[normalem]{ulem}
\usepackage{marvosym}
\usepackage{enumerate}

\definecolor{dodgerblue}{HTML}{1E90FF}
\definecolor{viennared}{HTML}{DA0A14}
\definecolor{ctorange}{HTML}{FF6C0C}
\definecolor{wales}{HTML}{ff0038}
\definecolor{benettongreen}{HTML}{009421}
\definecolor{ferrarired}{HTML}{ff2800}
\definecolor{austriawienpurple}{HTML}{441678}
\definecolor{chitot}{RGB}{255,127,14}
\definecolor{chip}{RGB}{148,103,189}
\definecolor{shockingpink}{HTML}{FF00D4}
\definecolor{skyblue}{RGB}{37,121,217}

\hypersetup{
     colorlinks=true,
     linkcolor=dodgerblue,
     filecolor=dodgerblue,
     citecolor = dodgerblue,      
     urlcolor=dodgerblue,
     }

\newcommand{\Birmingham}{School of Physics and Astronomy and Institute for Gravitational Wave Astronomy, University of Birmingham, Edgbaston, Birmingham, B15 2TT, United Kingdom}

\begin{document}

\title{Prospects for distinguishing dynamical tides in inspiralling binary neutron stars with third generation gravitational-wave detectors}

\author{Natalie Williams} 
\email{nwilliams@star.sr.bham.ac.uk}
\affiliation{\Birmingham}
\author{Geraint Pratten} 
\email{G.Pratten@bham.ac.uk}
\affiliation{\Birmingham}
\author{Patricia Schmidt} 
\email{P.Schmidt@bham.ac.uk}
\affiliation{\Birmingham}

\begin{abstract}
Tidal effects in gravitational-wave (GW) observations from binary neutron star mergers have the potential to probe ultra-dense matter and shed light on the unknown nuclear equation of state of neutron stars. Tidal effects in inspiralling neutron star binaries become relevant at GW frequencies of a few hundred Hz and require detectors with exquisite high-frequency sensitivity. Third generation GW detectors such as the Einstein Telescope or Cosmic Explorer will be particularly sensitive in this high-frequency regime, allowing us to probe neutron star tides beyond the adiabatic approximation. Here we assess whether dynamical tides can be measured from a neutron star inspiral. We find that the measurability of dynamical tides depends strongly on the neutron star mass and equation of state. For a semi-realistic population of 10,000 inspiralling binary neutron stars, we conservatively estimate that on average $\mathcal{O}(50)$ binaries will have measurable dynamical tides. 
As dynamical tides are characterised not only by the star's tidal deformability but also by its fundamental ($f$-) mode frequency, they present a possibility of probing higher-order tidal effects and test consistency with quasi-universal relations. For a GW170817-like signal in a third generation detector network, we find that the stars' $f$-mode frequencies can be measured to within a few hundred Hz. 
\end{abstract}

\date{\today}

\maketitle

\section{Introduction}
\label{sec:intro}
The detection of gravitational waves (GWs) from binary neutron star (BNS) mergers~\cite{TheLIGOScientific:2017qsa, LIGOScientific:2020aai} has opened up a new avenue to study the microscopic physics of neutron stars from their macroscopic properties. This inference is enabled by the imprint of tidal effects on the GW signal, which allows for the extraction of information about the as-of-yet unknown nuclear equation of state (EOS) of neutron stars~\cite{Flanagan:2007ix, Wade:2014vqa}.
The discovery of the BNS inspiral GW170817 allowed to place the first constraints on the nuclear EOS of neutron stars favouring a soft to medium-soft EOS with a median pressure at twice the nuclear saturation density of $3.5\times 10^{34}\, {\rm dyn\, cm^{-3}}$~\cite{GW170817-EOS}. In addition, complimentary constraints on the neutron star EOS have been obtained from the NICER observation of PSR J0740+6620~\cite{Miller:2021qha, Raaijmakers:2021uju}, and the neutron skin thickness of the lead isotope $\mathrm{{}^{208}Pb (R_{skin}^{208}})$ as measured by the PREX-2 experiment~\cite{Reed:2021nqk}.

Improvements in the sensitivity of the currently operating network of ground-based GW detectors Advanced LIGO~\cite{aLIGO}, Virgo~\cite{VIRGO:2014yos} and KAGRA~\cite{KAGRA:2020tym} will allow for observation of many more BNS signals~\cite{ObservingScenario} in the coming years. However, the third generation (3G) of GW detectors is particularly promising not only for the detection of all BNS out to a redshift of $z=3$~\cite{Maggiore:2019uih, Reitze:2019iox} but especially for precision measurements of the EOS from GW observations~\cite{Pacilio:2021jmq}.

The measurement of the neutron star EOS from BNS observations is made possible by the characteristic imprint left in the GW signal due to interaction of the star with its companion's gravitational field~\cite{Damour1983,Damour:1991yw,Flanagan:2007ix,Damour:2009vw,Vines:2011,Damour:2012yf}, which leads to the excitation of various fluid  oscillation modes. Here, we consider only the GW signature of the fundamental oscillation modes ($f$-modes) with $n=0$ radial nodes. The $\ell$-th multipolar oscillation mode is characterised by two parameters: The tidal deformability $\lambda_\ell$ and the angular mode frequency $\omega_\ell$. In the regime where $\omega_\ell$ is much smaller than the orbital frequency of the binary motion, the dominant tidal effects are \emph{adiabatic} $f$-modes, which depend purely on $\lambda_\ell$ and are known to 7.5 post-Newtonian (PN) order~\cite{Damour:2012yf, Narikawa:2021pak}. The observation of GW170817 allowed for the first measurement of the tidal deformability~\cite{TheLIGOScientific:2017qsa,Abbott:2018wiz,GW170817-EOS}. In the late inspiral at GW frequencies $\gtrsim 800$ Hz, finite-$\omega_\ell$ effects, referred to as \emph{dynamical} tides, become important, further enhancing the GW emission. The first constraints on the $f$-mode frequency of the companions of GW170817 were presented in~\cite{Pratten:2019sed}.

Assuming General Relativity (GR) and a hadronic composition of the neutron star, the tidal deformability and $f$-mode frequency 
can be related through quasi-universal relations (UR)~\cite{Chan:2014kua}. By directly measuring both dynamical and adiabatic tides, the assumptions behind these URs use can be tested, most prominently as a platform to test General Relativity and exotic matter models. Furthermore, the $f$-mode frequency is commonly contextualised as a property of the post-merger if a neutron star is formed~\cite{Andersson:1997rn, Stergioulas:2011gd, Bauswein:2011tp}, however, measuring it also during the inspiral would allow for consistency tests between the inspiral and remnant properties and to search for possible phase transitions during the merger. 

\begin{figure}
    \centering
    \includegraphics[width=\columnwidth]{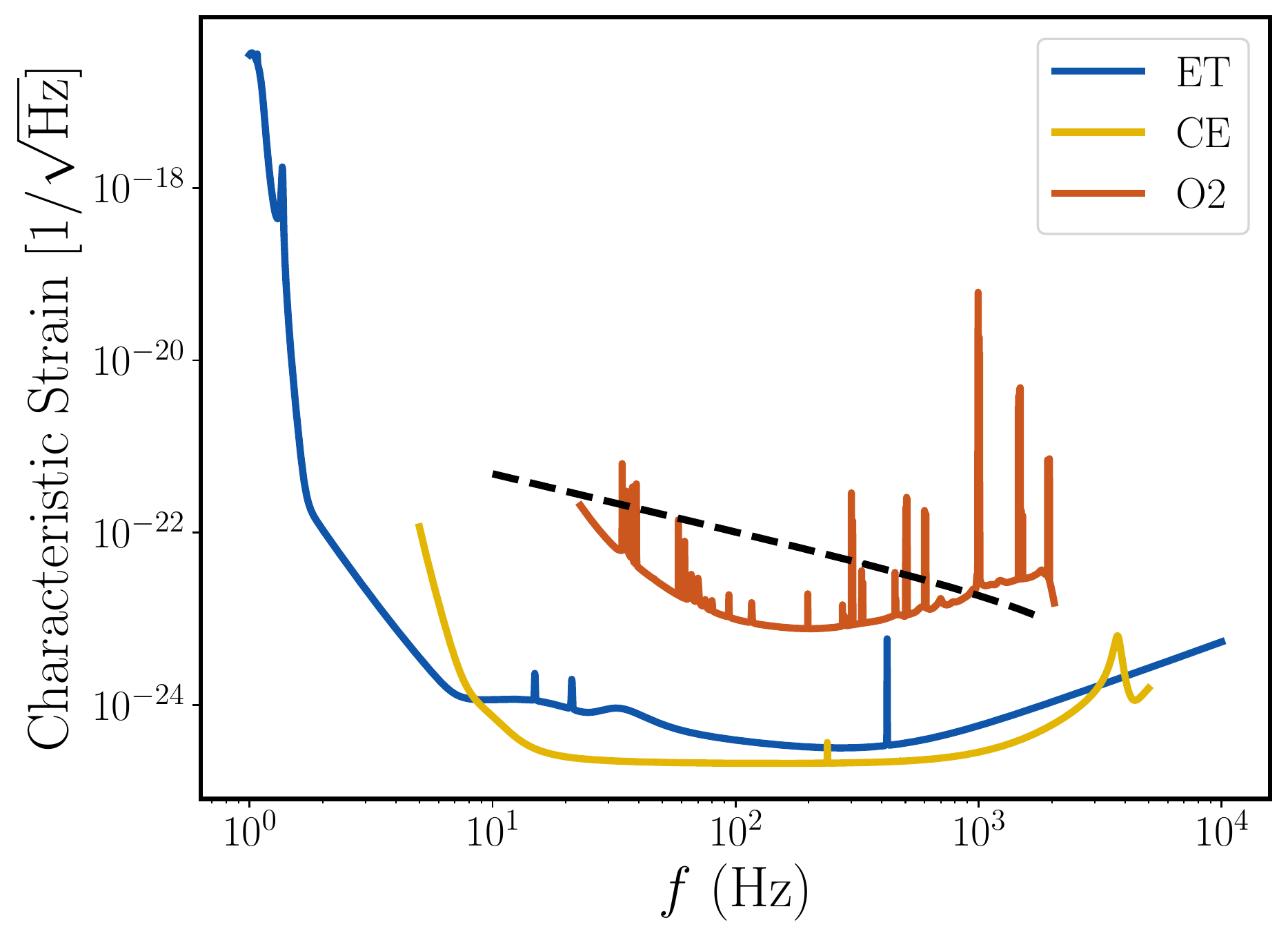}
    \caption{Sensitivity curves for ET (blue) and CE (yellow) ~\cite{PSD-P1600143} and the one for LIGO-Hanford at time of the observation of GW170817~\cite{PSD-P1900011} during the second observing run (O2) for comparison. The inspiral waveform for a  GW170817-like binary starting at $10$Hz is shown by the black dashed line. The regime above $\sim 800$Hz will be easily accessible for the 3G detectors.}
    \label{fig:PSDs}
\end{figure}

Current GW detectors are not sufficiently sensitive at frequencies above $\sim 800$Hz where dynamical tides become more prominent, and hence the measurements of the $f$-mode frequency is difficult. However, 3G detectors such as the Einstein Telescope (ET)~\cite{ET:2020aa} and Cosmic Explorer (CE)~\cite{Reitze:2019iox} will have a much improved sensitivity in the high-frequency regime as illustrated in Fig.~\ref{fig:PSDs}, and therefore the complete BNS signal through merger will be detectable for many of the anticipated $10^{4}$ detections per year \cite{Baibhav:2019gxm}, allowing us to also measure such higher-order tidal effects. 

In this paper we investigate the prospect for distinguishing dynamical tides in inspiralling BNS in 3G detectors. To do this, we calculate the signal-to-noise ratios (SNRs) required to distinguish between adiabatic and dynamical tides, and apply this to a fiducial BNS population to determine the proportion of events from which we can expect to measure dynamical tides. Furthermore, we perform full Bayesian inference on a GW170817-like binary to study how well the $f$-mode frequency can be measured.

The paper is organised as follows: In Sec.~\ref{sec:Method} we introduce the methodology including the waveform model we employ (Sec.~\ref{sec:waveform}), the distinguishability criterion (Sec~\ref{sec:distcriterion}), and provide a short summary of Bayesian inference in Sec.~\ref{sec:Bayesian}. In Sec.~\ref{sec:results} we present our results, first the distinguishability SNRs required to differentiate dynamical tides in Sec.~\ref{sec:rhodist} and then applied to realistic population of BNS Sec.~\ref{sec:Population}; in Sec.~\ref{sec:pe} we perform full Bayesian inference a GW170817-like event. Finally, we conclude in Sec.~\ref{sec:Conclusions}. Throughout we set $G=c=1$.

\section{Methodology}
\label{sec:Method}
\subsection{Waveform Model}
\label{sec:waveform}
To simulate the BNS, we model the frequency-domain GW phase $\Psi(f)$ of their signals as the post-Newtonian point-particle inspiral phase at 3.5PN order (see~\cite{Abbott:2018wiz,Pratten:2020fqn} and Refs. therein for details) augmented with quadrupolar ($\ell =2)$ adiabatic tidal effects up to 7.5PN order~\cite{Damour:2012yf, Vines:2011ud}, the 6PN octoplar $(\ell =3)$ adiabatic contribution~\cite{Hinderer:2009ca}, and the quadrupolar as well as octopolar dynamical tides contributions of Ref.~\cite{Schmidt:2019wrl}. For consistency with the dynamical tides prescription we only consider nonspinning neutron stars.\footnote{We note that recent work in the EOB framework has started to include the effect of the neutron star spin on the $f$-mode frequency and the associated GW phase \cite{Steinhoff:2021dsn}.} 
We do not include any tidal corrections in the PN amplitude. Hence, the BNS signal is schematically given as
\begin{equation}
    \tilde{h}_{\rm BNS}(f) = A_{\rm pp}(f) e^{i(\Psi_{\rm pp}(f) + \Psi_{\rm ad}(f) + \Psi_{\rm dyn}(f))},
\label{eq:waveform}
\end{equation}
where $A_{\rm pp}$ is the pure point-particle amplitude without tidal corrections, $\Psi_{\rm pp}$ the point particle contribution to the GW phase, $\Psi_{\rm ad}$ the adiabatic, and $\Psi_{\rm dyn}$ the dynamical phase all in the frequency domain.

The leading-order adiabatic tidal effects enters the GW phase at 5PN order and are entirely characterised by the binary tidal deformability $\tilde{\Lambda}$~\cite{Flanagan:2007ix, Wade:2014vqa}. The quadrupolar adiabatic tidal terms depend on the individual dimensionless tidal deformability of the A-th neutron star, $\Lambda_{2,A}=\lambda_{2,A}/m_A^5$, the octopolar term on $\Lambda_{3,A}=\lambda_{3,A}/m_A^7$. The dynamical tidal terms additionally depend on the stars' dimensionless angular $f$-mode frequencies $\Omega_{2,A}=m_A\omega_{2,A}$ and $\Omega_{3,A}=m_A\omega_{3,A}$. We consider these terms and neglect the quadrupole-monopole contribution~\cite{Poisson:1997ha} as we only consider nonspinning neutron stars.

For the majority of analyses we choose a waveform starting frequency of $f_{\rm min}=10$ Hz unless stated otherwise, and truncate the waveforms at a maximum frequency $f_{\rm max}$ either given by the (Newtonian) contact frequency~\cite{Agathos:2015uaa, Maselli:2013mva} or the frequency of the innermost stable circular orbit (ISCO) whichever one is smaller.

\subsection{Distinguishability Criterion}
\label{sec:distcriterion}
The agreement between two waveforms $h_1$ and $h_2$ is measured in terms of the match $\mathcal{M}(h_1,h_2)$ given by
\begin{equation}
\label{eq:match}
    \mathcal{M}(h_1,h_2) = \frac{\langle h_1|h_2\rangle}{\sqrt{\langle h_1|h_1\rangle \langle h_2|h_2\rangle}},
\end{equation}
where
\begin{equation}
    \langle h_1|h_2\rangle = 4 {\rm Re}\int_{f_{\rm min}}^{f_{\rm max}} \frac{\tilde{h}_1(f)\tilde{h}_2^*(f)}{S_n(f)} df,
    \label{eq:innerprod}
\end{equation}
is the noise-weighted inner product between two waveforms. Here, $S_n(f)$ is the PSD of the detector strain noise, $\tilde{h}(f)$ denotes the Fourier transform of $h(t)$ and ${}^*$ complex conjugation. The noise-weighted inner product of a GW $h$ with itself is the \emph{optimal} signal-to-noise (SNR) 
\begin{equation}
    \rho_{\rm opt}(h) = \sqrt{\langle h, h \rangle}.
\label{eq:rhotrue}
\end{equation}
We will use the PSDs of~\cite{PSD-P1600143} for the ET-D and CE sensitivities in our analysis. We place ET at the current Virgo site and CE at the current Hanford site of Advanced LIGO. 

Two waveforms $h_1$ and $h_2$ are considered \emph{indistinguishable} in a given detector if 
\begin{equation}
    1-\mathcal{M}(h_1, h_2) < \frac{D}{2\rho^2},
\label{eq:dist}
\end{equation}
where $\rho$ is the \emph{measured} SNR and $D$ denotes the number of measurable intrinsic binary parameters~\cite{Chatziioannou:2017tdw, Lindblom:2008cm}. Conversely, two signals can be distinguished if the measured SNR is larger than the necessary \emph{distinguishability} SNR defined as
\begin{equation}
    \rho_{\rm dist} = \sqrt{\frac{D}{2(1-\mathcal{M})}}.
\label{eq:rhodist}
\end{equation}
We will use the distinguishability SNR in Sec.~\ref{sec:rhodist} to determine where in the BNS parameter space dynamical tides will be, in principle, measurable in 3G detectors.

\subsection{Bayesian Inference}
\label{sec:Bayesian}
While the distinguishability criterion is a sufficient condition to gauge for which binaries dynamical tides may be significant, it does not allow us to determine to what degree the $f$-mode frequencies $\Omega_{\ell, A}$ can be constrained in future BNS observations. Therefore, we perform full Bayesian inference on a 3G detector networks for a GW170817-like BNS merger.
Bayes theorem states that the posterior density function (PDF) of a set of parameters $\boldsymbol\theta$ given the data $d$ is
\begin{equation}
    p(\boldsymbol\theta| d) = \frac{p(d|\boldsymbol\theta) p(\boldsymbol\theta)}{p(d)},
\end{equation}
where $p(d|\boldsymbol\theta)$ is the likelihood, $p(\boldsymbol\theta)$ the prior reflecting any prior knowledge about the parameters, and $p(d)$ is the evidence or marginalised likelihood,
\begin{equation}
    p(d) =\int p(d|\boldsymbol\theta) p(\boldsymbol\theta) d\boldsymbol\theta,
\end{equation}
which serves as a normalisation factor. 

For the BNS systems that we will consider in Sec.~\ref{sec:pe}, $\boldsymbol\theta$ consists of the intrinsic parameters of the binary, i.e. the component masses $m_A$, the quadrupolar tidal deformabilities $\Lambda_{2,A}$ and the quadrupolar $f$-mode frequencies $\Omega_{2,A}$ (all octopolar contributions are omitted) with $A={1,2}$, and extrinsic parameters, i.e. the sky location, inclination, distance, polarisation, coalescence time and phase.

For a detector network the joint likelihood is given by the product of the individual detector likelihoods, 
\begin{equation}
    p(d|\boldsymbol\theta) =\prod_{i=1}^{N} p(d_i|\boldsymbol\theta),
\end{equation}
where $N$ is the number of detectors in the network. 

One-dimensional and two-dimensional PDFs are obtained by marginalising over the other parameters. We utilise the \textsc{Bilby} inference library~\cite{Ashton:2018jfp, Romero-Shaw:2020owr} in conjunction with the nested sampler \textsc{dynesty}~\cite{Speagle:2020aa} using slice sampling to sample the parameter space and estimate the posterior density. We inject simulated signals into zero noise, which is broadly equivalent to the results obtained by averaging the PSD over many noise realisations. 
However, we do not neglect noise entirely as it still enters the likelihood calculation through the PSD.

\section{Results}
\label{sec:results}
\subsection{Distinguishability SNR for BNS}
\label{sec:rhodist}

\begin{figure*}
    \centering
    \includegraphics[width=\textwidth]{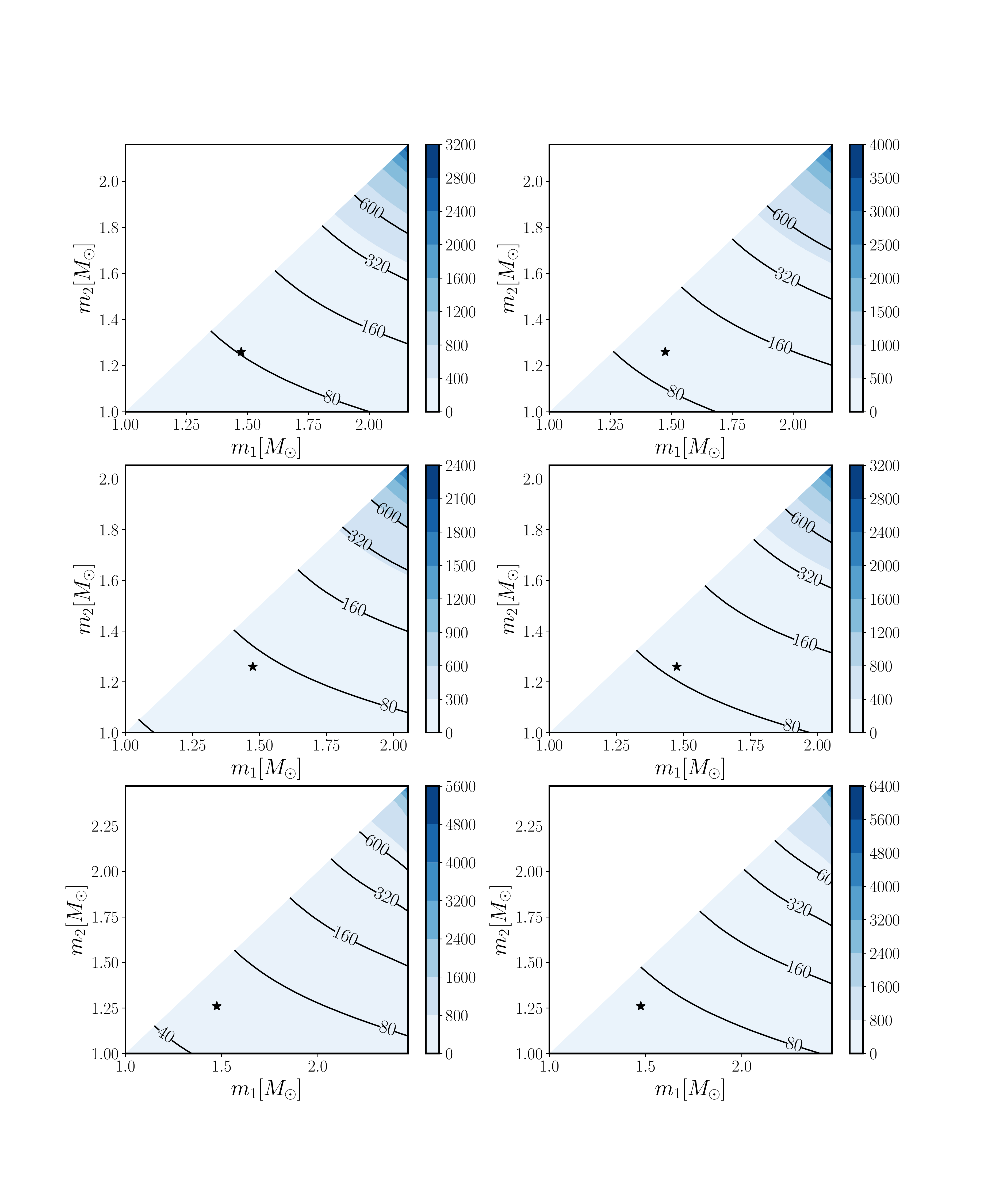}
    \caption{Contour plots for the minimal SNR required to distinguish between adiabatic and dynamical tides as a function of component mass $m_1, m_2$ for ET (left column) and CE (right column). We show results for three different EOS: APR4 (top row), SLy (middle row) and MPA1 (bottom row). The black star represents a GW170817-like event.}
    \label{fig:distinguishability}
\end{figure*}

Despite their high relative PN order, dynamical tides are expected to be distinguishable from adiabatic tides in the late inspiral of BNS if the SNR is large enough. Here, we perform a first assessment of the SNR required to separate adiabatic and dynamical tides in 3G detectors for a range of neutron star masses. Considering only nonspinning BNS with tides up to octopolar order, we compute the distinguishability SNR $\rho_{\rm DT}$ following Eq.~\eqref{eq:rhodist} with $D=6$, where the six intrinsic parameters are $m_A, \Lambda_{2,A}, \Omega_{2,A}$\footnote{We do not include the $\ell=3$ tidal parameters in $D$ as these are even more difficult to measure but their neglect in waveform models may induce a bias in the $\ell=2$ tidal parameters and hence they are included in our inspiral model.} and the match $\mathcal{M}$ between a waveform containing only adiabatic tides, $h_{\rm ad}$, and one that includes both adiabatic and dynamic tides, $h_{\rm dyn}$, for the same parameters for ET and CE with $f_{\rm min} = 10$ Hz. Both waveforms $h_{\rm ad}$ and $h_{\rm dyn}$ include quadrupolar and octopolar adiabatic tidal terms; $h_{dyn}$ additionally includes the quadrupolar and octopolar dynamical tidal terms.

We consider three hadronic EOS representative of soft to medium-soft EOS consistent with GW170817~\cite{GW170817-EOS}: APR4~\cite{Akmal:1998cf}, SLy~\cite{Reinhard:1995rf,Danielewicz:2008cm,Gulminelli:2015csa} and MPA1~\cite{Muther:1987mp} and use the UR of Ref.~\cite{Chan:2014kua} to evaluate the quadrupolar $f_2$-mode and octopolar $f_3$-mode frequencies. In reality the EOS is unlikely to adhere to any specific model listed here, but lie somewhere in the parameter space that they cover. We choose a minimum NS mass of $1\, M_\odot$ up to the maximum mass allowed by each EOS. Figure~\ref{fig:distinguishability} shows the distinguishability SNR for dynamical tides for ET (left column) and CE (right column) for APR4 (top row), SLy (middle row) and MPA1 (bottom row). In all panels we indicate a GW170817-like binary with mass ratio $q_\star = 1.17$ and source-frame total mass $M^{\rm src}_\star = 2.735\, M_\odot$ with a star.

\begin{figure}
    \centering
    \includegraphics[width=\columnwidth]{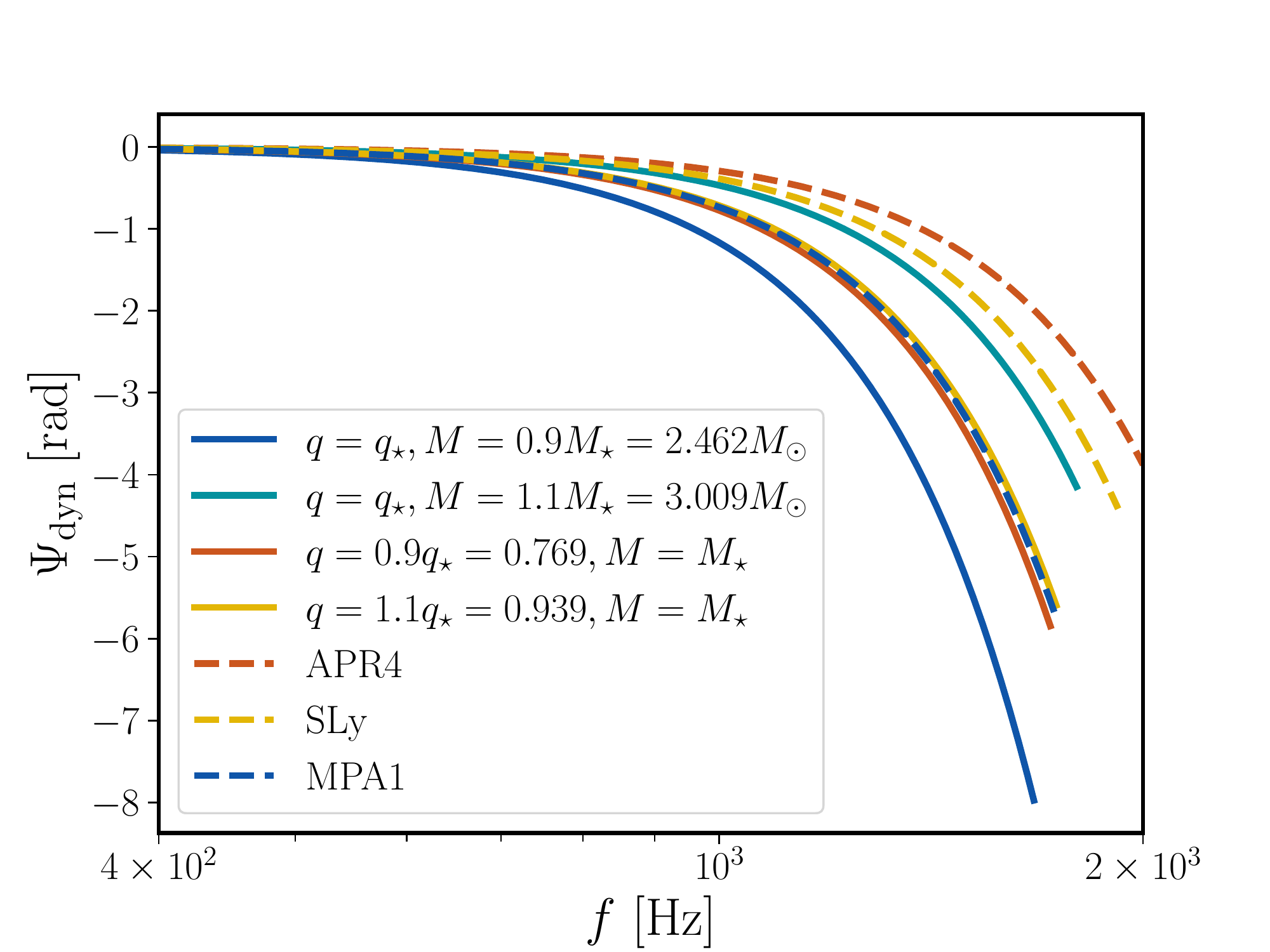}
    \caption{GW phase contribution from dynamical tides, $\Psi_{\rm dyn}$, as a function of GW frequency for a variation of total mass and/or mass ratio (solid lines) and EOS (dashed lines). We either vary the total mass $M$ or the mass ratio $q$ by $10$\% from a GW170817-like binary with $q_{\star} = 0.854$, and $M_{\star}=2.735 M_{\odot}$ and EOS MPA1. We also vary the EOS for a GW170817-like binary showing the tidal phase for APR4 (orange), SLy (yellow) and MPA1 (blue).}
    \label{fig:dephasing}
\end{figure}

The dependence of $\rho_{\rm DT}$ on the total mass $M=m_1 +m_2$ and EOS is a consequence of their effect on $\Psi_{\rm dyn}$. Figure \ref{fig:dephasing} shows the dynamical tides part of the phase for 10\% changes in either the total mass or mass ratio $q = m_1/m_2 \geq 1$ for a GW170817-like system. It is evident that $\Psi_{\rm dyn}$ is maximised for low total masses. Consequently, we see that for heavy BNS systems the dynamical tides are suppressed and therefore more difficult to measure. Variation in EOS shows that for a given pair of $(M, q)$, stiffer EOS (e.g. MPA1) produce the largest contribution to the dynamical tidal phase for the EOS considered here. Physically, this corresponds to between a quarter and half an orbit dephasing at the contact frequency solely due to dynamical tides for our softest and stiffest EOS in the case of GW170817.
These observations explain the distribution of $\rho_{\rm DT}$ in Fig. \ref{fig:distinguishability}: At regions where $\Psi_{\rm dyn}$ is maximised, i.e. lighter neutron stars, a lower $\rho_{\rm DT}$ is required to disentangle adiabatic and dynamical tides. The SNRs required to see a noticeable impact of dynamical tides on the tidal phase reach up to $6400$ in ET for the MPA1 EOS, which admits the largest neutron star mass of the EOS considered here. 
However, when quantifying these SNRs, the overall increased sensitivity of 3G detectors must be taken into account as illustrated below.

Considering a GW170817-like event as a typical event, we measure $\rho_{\rm DT}$ = 84, 74, 56 (102, 89, 67) for APR4, SLy and MPA1 respectively in ET (CE).
For comparison, the measured SNR of GW170817 in the LIGO-Virgo network was $\sim 32$~\cite{TheLIGOScientific:2017qsa}. For ET and CE the optimal SNRs for a GW170817-like binary would be 1031 and 2935 respectively for this event. In this scenario, any EOS considered here would lead to results that exceed the minimum distinguishability SNR $\rho_{\rm DT}$ by at least a factor of 10. It is also important to note that these results consider only one detector, while for any coincident event the network SNR is the quadrature sum of the individual detector SNRs.

\begin{figure*}
    \centering
    \includegraphics[width=\textwidth]{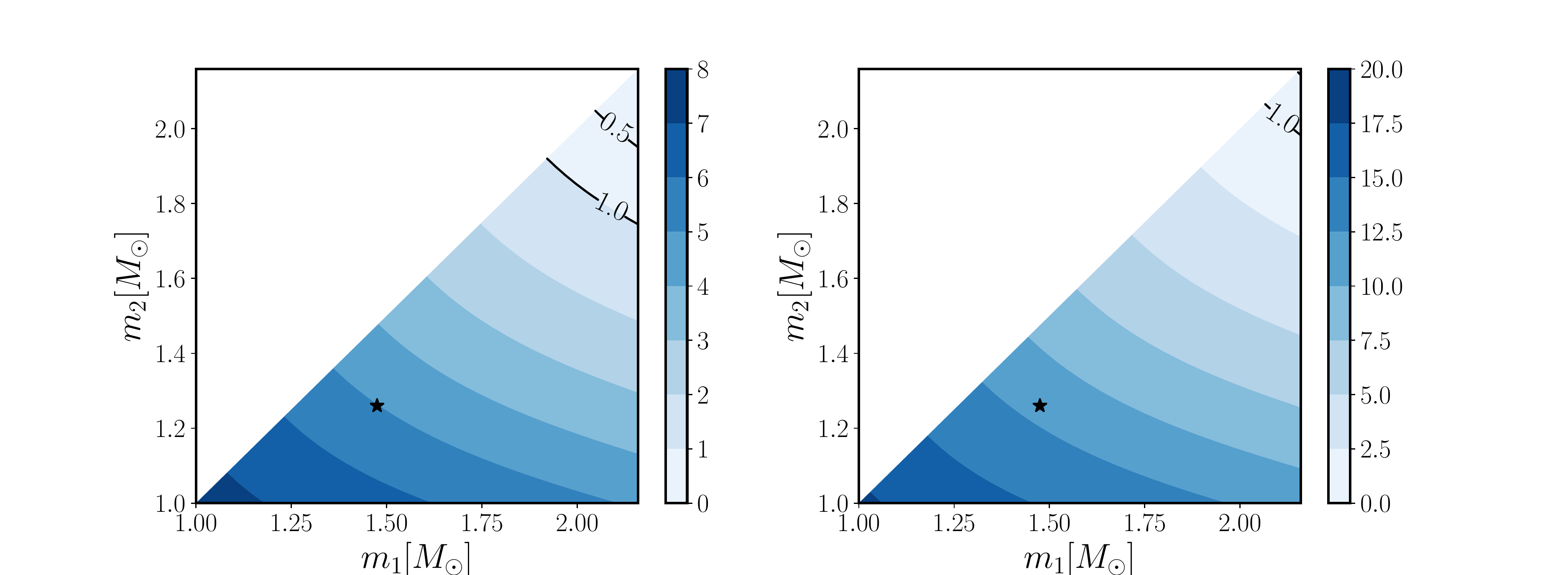}
    \caption{Contour plots for $\rho_{\rm opt}/\rho_{\rm DT}$ for an optimally orientated source at 100Mpc with EOS APR4, as a function of component mass $m_1, m_2$ for ET (left) and CE (right). The black star represents a GW170817-like event. Dynamical tides are distinguishable when $\rho_{\rm opt}/\rho_{\rm DT} > 1$.}
    \label{fig:distinguishabilityratio}
\end{figure*}

When comparing the two 3G detectors, we consistently find that CE requires larger values of $\rho_{\rm DT}$. It may seem counterintuitive to require a higher distinguishability SNR for a more sensitive detector at first glance, however, the higher optimal SNR (Eq.~\eqref{eq:rhotrue}) of a signal detected by CE must be remembered comparatively. Therefore, to meaningfully compare the two 3G detectors, it is necessary to also take the optimal SNR into account. 
We demonstrate a comparison between detectors in Fig. \ref{fig:distinguishabilityratio}, where we mirror Fig. \ref{fig:distinguishability} but show the ratio $\rho_{\rm opt}/\rho_{\rm DT}$ assuming the BNS are at a luminosity distance of $100$ Mpc. 
This ratio can be interpreted as the fraction of the optimal SNR that is required to distinguish between adiabatic and dynamical tides. It follows that for $\rho_{\rm opt}/\rho_{\rm DT} > 1$, it is possible to distinguish dynamical tides in this signal for the chosen luminosity distance. Therefore despite the higher $\rho_{\rm DT}$ in CE relative to ET, the  required fraction of the optimal SNR is much lower, and hence dynamical tides are more easily detectable. For example, in the case of a GW170817-like system at $100$ Mpc $\rho_{\rm DT} = \rho_{\rm opt}/5$ for ET and $\rho_{\rm DT} = \rho_{\rm opt}/11$ for CE.

We also explored the distinguishability of octopolar ($\ell=3$) dynamical tides from quadrupolar dynamical tides, which, as expected, requires even higher SNRs. As reference, for a GW170817-like binary the SNRs required to discriminate the two would be 1258, 1050, 684 (1545, 1278, 822) for APR4, SLy and  MPA1 respectively in ET (CE). Comparing this to the optimal SNRs of 1031 (ET) and 2935 (CE), it is evident that for the softer EOS (as preferred by GW170817) ET is unlikely to be able to disentangle octopolar from quadrupolar dynamical tides, whereas CE would be able due to its increased sensitivity.

\subsection{Population Study}
\label{sec:Population}
We now consider a realistic population of BNS to which we apply the distinguishability methodology for our choices of EOS. 
We assume that the EOS is universal, i.e. it is the same for all NS.

Following the neutron star population model outlined in Ref.~\cite{Galaudage:2020zst}, we sample NS masses from the formation channel corresponding to the second-born NS of the binary system, referred to as ``slow'' neutron stars due to their effectively zero spin. Ref.~\cite{Galaudage:2020zst} assumes here that the first and second-born mass distributions are independent. This population takes the form of a double Gaussian in which the probability of a neutron star with mass $m$ is given by
\begin{equation}
    p(m) = \xi_s\mathcal{N}(\mu_1, \sigma_1) + (1 - \xi_s)\mathcal{N}(\mu_2, \sigma_2),
\end{equation}
where $\mathcal{N}(\mu_i, \sigma_i)$ are normalised Gaussians with mean $\mu_i$ and standard deviation $\sigma_i$, and $\xi_s$ is the fraction of binaries in the low mass peak. The values for $\xi_s, \mu_1, \sigma_1, \mu_2, \sigma_2$ are taken from Tab. 1 of Ref.~\cite{Galaudage:2020zst}. Here the low mass peak corresponds to radio-visible NS, and the high mass peak is informed by GW190425 as originating from a fast merging population.

Although the BNS redshift horizon of 3G detectors is predicted to reach up to $z \sim 3$ \cite{Maggiore:2019uih,Reitze:2019iox}, we recall from Sec. \ref{sec:distcriterion} that high SNRs are required to distinguish higher order tidal effects, and we therefore introduce a redshift cut-off of $z = 0.5$ (equivalent to a luminosity distance $D_L \sim 3000$ Mpc assuming a Planck15~\cite{Planck:2015fie} cosmology) for our population study. We also impose a minimum SNR of $\rho_{\rm opt}> 8$ for each detector. Considering a GW170817-like event placed at $z=3$, the measured SNR would be $\sim 5$ (6) for ET (CE), which would not be detectable and demonstrates our choice for the distance cut. We distribute the BNS in redshift following the Madau-Dickinson star formation rate (SFR)~\cite{Madau:2014bja}, as at low redshift regardless of the time delay distribution chosen between birth and merger, merger rates broadly follow the SFR. We specifically adopt the distribution described in Ref.~\cite{Fishbach:2018edt}, in which the probability distribution of redshift $z$ is given by
\begin{equation}
    p(z) = \frac{dV_c}{dz} \frac{1}{1+z} \psi(z), 
\end{equation}
where $V_c$ is the comoving volume and $\psi (z)$ is the SFR 
\begin{equation}
    \psi (z) = 0.015 \frac{(1+z)^{2.7}}{1 +[(1+z)/2.9]^{5.6}} M_\odot \textrm{yr}^{-1} \textrm{Mpc}^{-3}.
\end{equation}
Current GW observations constrain the local merger rate for BNS to be $\sim 10-1700 \, \rm{Gpc}^{-3} \, \rm{yr}^{-1}$ \cite{LIGOScientific:2021psn}. Following \cite{Finn:1992xs}, we estimate the detection rate of BNS up to $z \sim 0.5$ for a single CE detector to be $\sim \mathcal{O}(\rm{few} \times 10^4) \, \rm{yr}^{-1}$, which is consistent with the estimated detection rate for BNS mergers in 3G detectors ~\cite{Baibhav:2019gxm}. We therefore generate ten (random) realisations of $10^4$ binaries, where the gpstime of each binary is chosen randomly from the uniform interval $[1703721618, 1735257618]$, to place a conservative bound on the observability of dynamical tides from a population of BNS.

Extrinsic parameters are randomly drawn from uniform distributions, where declination $\delta$ is sampled in cosine, inclination $\iota$ in sine, and right ascension $\alpha$ and polarisation angle $\psi$ are sampled between 0 and $\pi$. The optimal SNR $\rho_{\rm opt}$ is then calculated as in Eq.~\eqref{eq:rhotrue}, however, due to varying extrinsic parameters there now also exists a geometric factor folded in via the detector response.

\begin{table}[h]
\begin{tabular}{c|c|c|c|c|}
\cline{2-5}
& \multicolumn{2}{c|}{$\rho_{\rm opt}\geq\rho_{\rm AT}$} & \multicolumn{2}{c|}{$\rho_{\rm opt}\geq\rho_{\rm DT}$} \\ 
\cline{2-5} 
& \rule{0pt}{2.5ex}  ET & CE & ET & CE   \\ 
\hline
\multicolumn{1}{|c|}{APR4} & 404.1  & 1018.4 & 14.9 &  43.0 \\
\multicolumn{1}{|c|}{SLy}  & 558.1 & 1309.8 & 22.3 &  68.6  \\
\multicolumn{1}{|c|}{MPA1} & 1150.5 & 2392.6 & 46.8 &  149.7 \\ 
\hline
\end{tabular}
\caption{The average number of events detected by ET and CE where adiabatic tides are distinguishable $\rho_{\rm opt}\geq \rho_{\rm AT}$, and the subset of these for which also dynamical tides are distinguishable $\rho_{\rm opt}\geq \rho_{\rm DT}$ from 10 realisations of $10^4$ BNS for different EOS.}
\label{table:number_dist}
\end{table}

\begin{figure*}
    \centering
    \includegraphics[width=\textwidth]{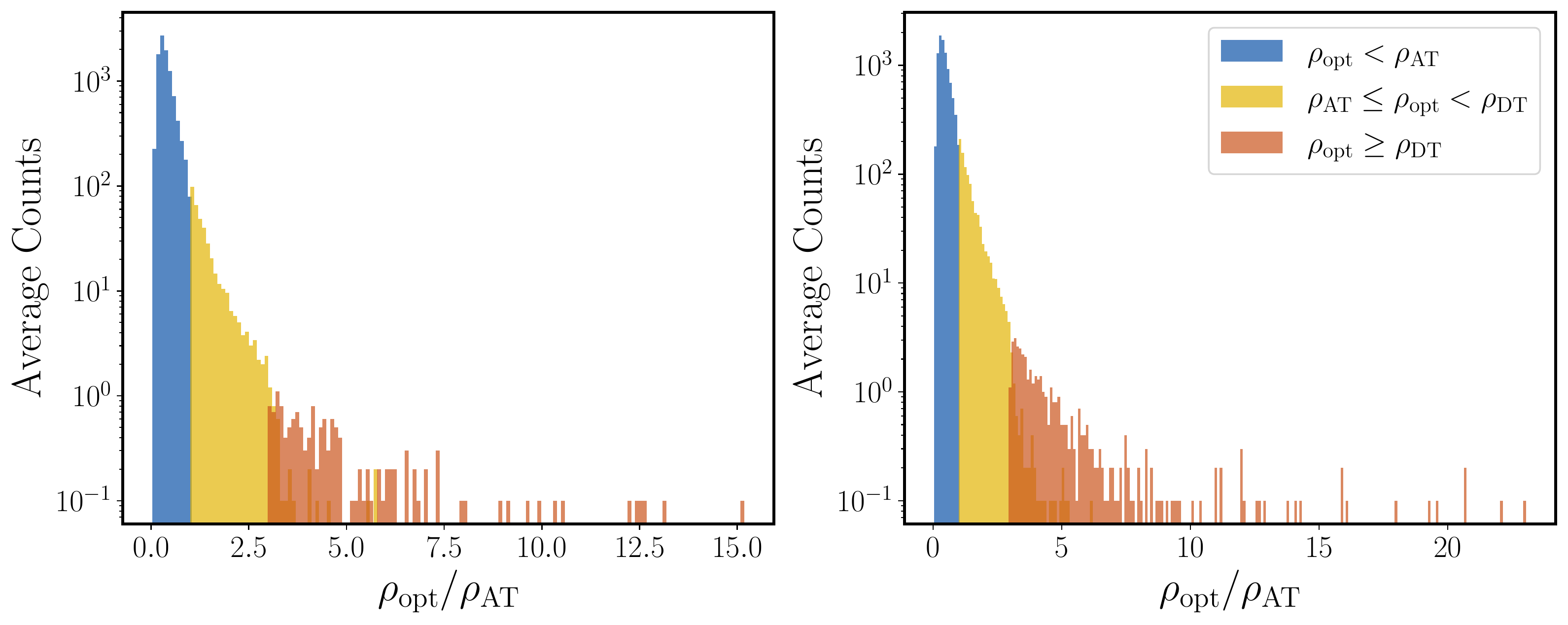}
    \caption{Histograms of the average number of BNS events as a function of the SNR divided by the distinguishability SNR for adiabatic tides for ET (left) and CE (right) for APR4. All events with $\rho_{\rm opt}/\rho_{\rm AT} < 1$ (blue) are indistinguishable from binary black hole signals; for events with  $\rho_{\rm AT}\leq \rho_{\rm opt}< \rho_{\rm DT}$ (yellow) the adiabatic contribution to the phase can be distinguished; for events that satisfy $\rho_{\rm opt}\geq \rho_{\rm DT}$ (orange) both adiabatic and dynamical contributions to the phase are distinguishable.}
    \label{fig:disthist}
\end{figure*}

Table \ref{table:number_dist} shows the number of events for which tidal information is recovered when averaged over our ten realisations, i.e. all binaries satisfying $\rho_{\rm opt}\geq \rho_{\rm AT}$, where $\rho_{\rm AT}$ is the distinguishability SNR Eq.~\eqref{eq:dist} for adiabatic tides. 
No tidal information would be recovered in cases that do not satisfy this, making them indistinguishable from binary black hole events, and the only evidence for a BNS would be from the component masses. We also list the subset of events that have detectable dynamical tides, i.e. where $(\rho_{\rm opt}\geq \rho_{\rm DT})$. 
Considering the stiff EOS MPA1, tidal information is measurable in 11.5\% (23.9\%) of binaries, and for 4.2\% (6.7\%) of those also dynamical tides can be recovered in ET (CE). This fraction decreases for softer EOS such as APR4, reducing to 4.0\% (10.2\%) of events with measurable tides, and 3.7\% (4.2\%) of those have distinguishable dynamical tides. Figure \ref{fig:disthist} shows the average number of BNS as a function of $\rho_{\rm opt}/ \rho_{\rm AT}$ for the soft APR4 EOS. For events with $\rho_{\rm opt}/ \rho_{\rm AT} < 1$, no tidal information is measurable.
For all EOS considered here, the vast majority of events would not be distinguishable from BBH events. However, taking into account that the predicted BNS detection rates for a network of 3G detectors is on the order of $\sim 10^3 - 10^5$ binaries per year, we still expect a substantial number of binaries to have dynamical tidal contributions depending on the true EOS and detector network. From Table~\ref{table:number_dist}, we expect $\sim \mathcal{O}(\rm{few} \times 10^3)$ binaries per year with distinguishable adiabatic tides and $\sim \mathcal{O}(50)$ binaries per year with distinguishable dynamical tides with a single CE detector.

\subsection{Parameter Estimation}
\label{sec:pe}
Whilst the distinguishability criterion Eq.~\eqref{eq:dist} is a useful measure to estimate for which events dynamical tides may be significant, it does not inform on the accuracy to which the $f_2$-mode frequency can be constrained from a GW observation. 
Thus, we perform full Bayesian inference on a GW170817-like BNS signal in an ET-CE 3G detector network to determine the posterior probability densities of the tidal parameters and, in particular, the accuracy to which the $f_2$-mode frequency can be measured during the inspiral. Here, we only consider quadrupolar adiabatic and dynamical tides. We use the same waveform model for our simulated GW signal and to infer the parameters as detailed in Sec. \ref{sec:waveform}. 

\begin{figure*}
    \centering
    \includegraphics[width=\textwidth]{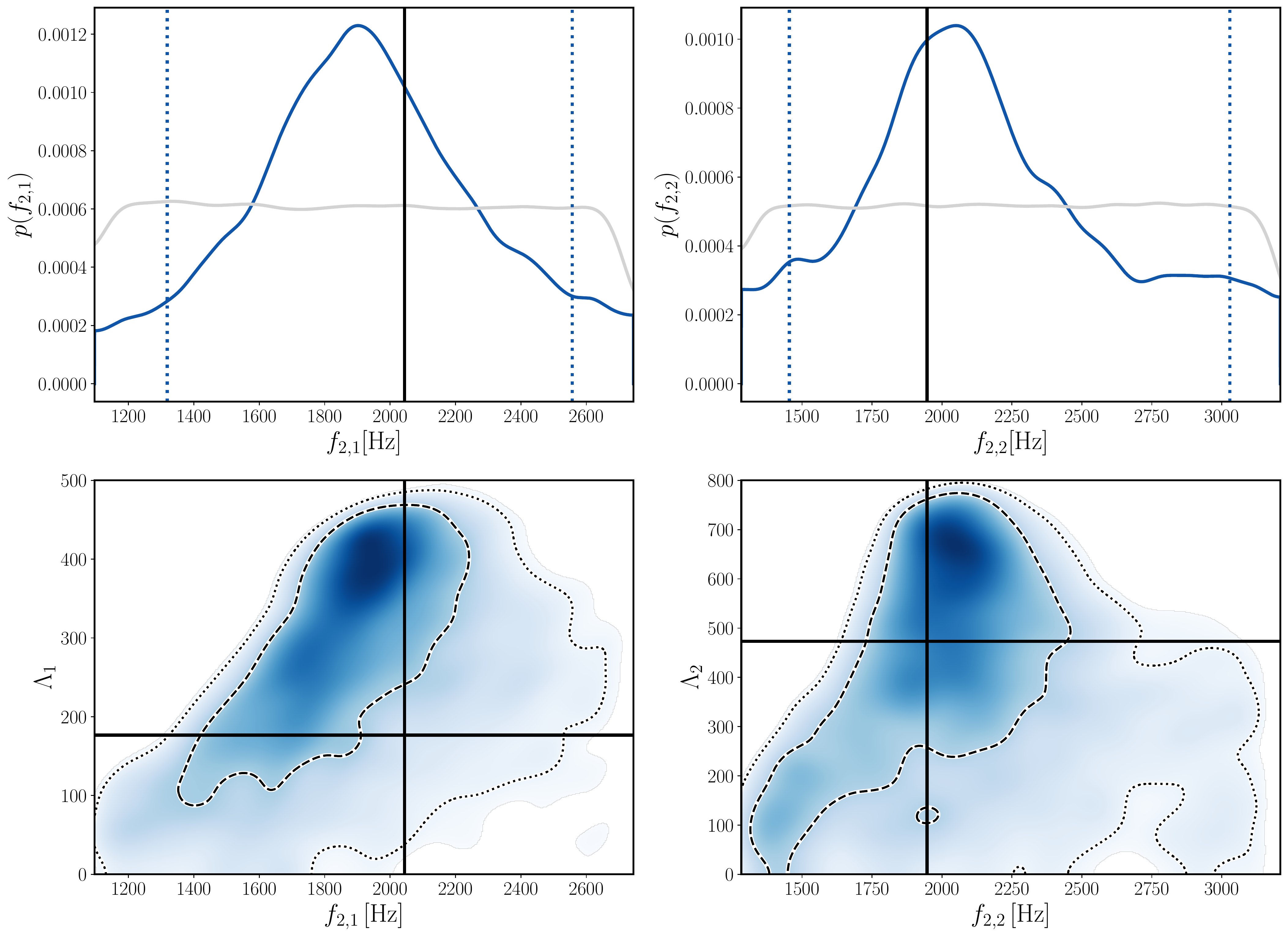}
    \caption{Posterior distributions for tidal parameters. Top: 1D posterior of the quadrupolar $f$-mode frequencies $f_{2,A}$ for the primary (left) and the secondary (right). Bottom: Joint 2D posteriors of $f_{2,A}$ and the tidal deformability $\Lambda_{A}$. The injected values (black solid lines) are shown alongside 50\% (dashed lines), 90\% (dotted lines) credible intervals/contours and priors (grey lines).}
    \label{fig:posteriors}
\end{figure*}

Consistent with GW170817 we choose source-frame component masses of the primary $m_1=1.475\, M_\odot$ and the secondary $m_2 = 1.26\, M_\odot$, and the soft APR4 EOS for the simulated signal. The extrinsic parameters are as follows: The inclination between the orbital angular momentum and the line-of-sight $\iota =0.1$ rad, right ascension $\alpha = 2.554$ rad, declination $\delta = -0.41$ rad, luminosity distance $D_L = 40$Mpc. We set the phase $\phi$, polarisation angle $\psi$ and GPS time to zero.

To reduce the computational cost of the sampling, we assume that the sky location $(\alpha, \delta)$, distance $d_L$ and polarisation $\psi$ are known. This assumption is justified by the detection of an electromagnetic counterpart to pin the sky location and distance, which, for a nonspinning system, trivially gives a measurement of the polarisation.\footnote{We note that electromagnetic counterparts are expected to be observable up to distances of $\sim 200$ Mpc for the next generation of telescopes~\cite{LSST:2018bbx}.} 
We integrate the likelihood from a GW frequency $f_{\rm min} = 20$ Hz up to the Schwarzschild ISCO frequency of $1593$ Hz. We note that for the chosen parameters, the BNS contact frequency is estimated to be $1843$ Hz, which is significantly higher than the point at which we truncate the analysis and hence it is safe to assume that the two stars are still well separated and that the $f_2$-mode frequency of each neutron star is a well defined quantity.
While the choice of the lower frequency cutoff reduces the signal length and hence the accumulated SNR, (adiabatic) tidal contributions to the phase only become prominent at GW frequencies $\gtrsim 400$ Hz~\cite{Flanagan:2007ix} and dynamical tides become relevant at even higher frequencies~\cite{Schmidt:2019wrl}. We therefore expect the impact of this choice to be small for the measurement of the $f_2$-mode frequency. Our simulated signal has an ET-CE network SNR of 2360 which per the results of Sec.~\ref{sec:distcriterion} easily satisfies the distinguishability criterion for a GW170817-like binary and hence we expect that some information about the $f_2$-mode 
frequency can be recovered from such an observation.

We obtain posterior probability distributions of the source parameters using the Nested Sampling algorithm~\textsc{Dynesty}~\cite{Speagle:2020aa}, as implemented in \textsc{Bilby}~\cite{Ashton:2018jfp}, with random slice sampling and $2000$ live points. The simulated GW signal is injected into zero noise to avoid any noise-induced biases in the results.
For our fiducial analysis we sample in chirp mass $\mathcal{M}_c$ and the inverse mass ratio $1/q$ and adopt the following uniform priors: $\mathcal{M}_c^{\rm det}\in[1.19666, 1.19675] M_\odot$, $1/q \in [0.825,0.875]$, $\Lambda_{2,A} \in [0,1000]$, $\Omega_{2,A}\in [0.05, 0.125]$, phase $\phi \in [0, 2\pi]$ and geocentric time $ [-0.1, 0.1]$ around the injected value. 

Figure~\ref{fig:posteriors} shows the 1D and 2D posteriors of the tidal parameters. The complete results are shown in Fig.~\ref{fig:full_run} in Appendix~\ref{sec:app}. At 90\% confidence we find the median $f$-mode frequencies of the stars to be $f_{2,1} = 1916^{+641}_{-598}$ Hz, and $f_{2,1} = 2103 ^{+926}_{-649}$ Hz, compared to the injected values of $f_{2,1} = 2044$ Hz and $f_{2,2} = 1947$ Hz. While the recovered median values are in good agreement with the injected values, the 90\% credible intervals are wider than $1000$ Hz despite the large SNR. From the 2D posteriors it becomes evident that there is a noticeable correlation between the $f_2$-mode frequency and $\Lambda_2$. As an attempt to break this correlation, we repeat the inference but choose a different mass prior. This choice is motivated by examination of the quadrupolar contribution of the dynamical phase, which is dependent on various combinations of component masses and tidal parameters. Noting the dependence of the dimensionless tidal deformability on the mass, by improving the component mass measurement we expect to improve constraints on the tidal parameters, leading to an improvement in the measurement of $\Omega_A$. To do so, we first determine the component mass posteriors from a low-frequency analysis from $10-20$ Hz entirely without tides and use the resulting mass posterior distributions as prior for the subsequent tidal analysis. For this restricted analysis, the SNR is 1582, and we sample again in $\mathcal{M}_c$ and $q$, keeping the $\mathcal{M}_c$ prior as before, however widening the mass ratio uniform prior to $q \in [0.5, 1]$. We also restrict the allowed range for the component masses to be $m_A \in [1,3]M_\odot$.
Everything else remains unchanged relative to our fiducial analysis. Once the mass posteriors are determine from the restricted analysis, we proceed with the tidal analysis from 20Hz as before but now sample directly in the component masses. The mean and variance are $\mu_1=1.4465, \sigma_1=0.0320$ and $\mu_2=1.3078, \sigma_2=0.0285$, respectively. 

\begin{figure*}
    \centering
    \includegraphics[width=\textwidth]{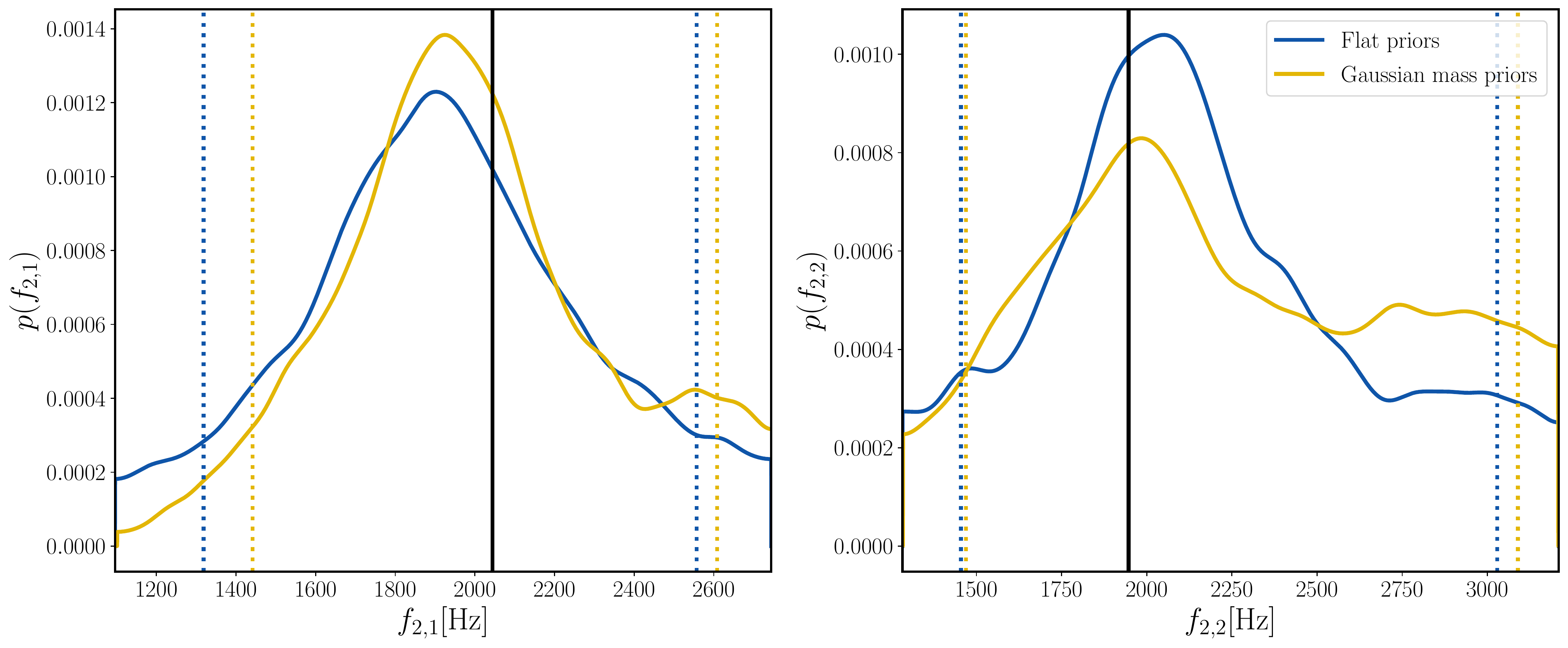}
    \caption{One-dimensional posterior distributions of the quadrupolar $f$-mode frequency for the primary (left) and secondary (right) neutron star for a GW170817-like binary in an ET-CE detector network. Flat priors (blue) and Gaussian component mass priors (yellow) are shown alongside the injected values (black solid lines) and the 90\% credible intervals (dotted lines).}
    \label{fig:gaussian_posteriors}
\end{figure*}

Figure~\ref{fig:gaussian_posteriors} shows the resulting 1D posteriors of $f_{2,A}$ in comparison to the results with the uniform mass prior. With the Gaussian mass priors we find the $f_2$-mode frequencies at 90\% confidence to be $f_{2,1} = 1959^{+649}_{-517}$ Hz, and $f_{2,2} = 2154^{+935}_{-684}$ Hz. 
When considering the primary mass, the Gaussian mass prior results show improvement in the posterior around the injected value and in turn a reduction in the 90\% interval. However, in the case of the secondary mass, no improvement is evident. This is unsurprising, as tidal deformation is enhanced for smaller masses, and thus the regime in which we can gain the most improvement is for the larger primary mass. The complete results for the analysis with the Gaussian mass prior and the non-tidal $10-20$ Hz run are shown in Figs.~\ref{fig:gaussian_run} and~\ref{fig:10_20} in Appendix~\ref{sec:app}.

\begin{figure*}
    \centering
    \includegraphics[width=\textwidth]{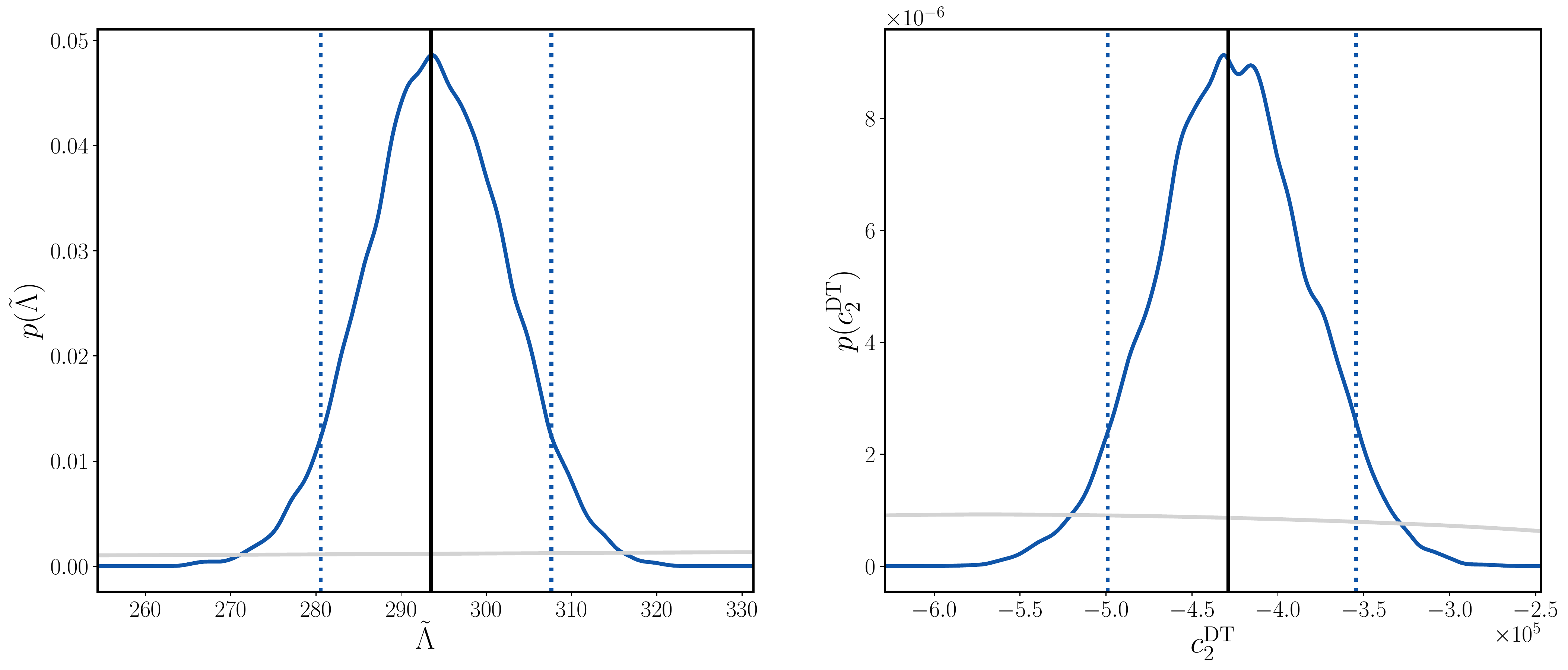}
    \caption{\emph{Left}: One-dimensional posterior distribution for $\tilde{\Lambda}$ for the fiducial run, showing the 90\% credible intervals (dotted lines), injected value (solid black line) and prior (solid grey line) with prior range $\tilde{\Lambda} \in [0, 1025]$. \emph{Right}: One-dimensional posterior distribution for $c_2^{\rm DT}$ for the fiducial run, showing the 90\% credible intervals (dotted lines), injected value (solid black line) and part of the prior (solid grey line) with the full prior range spanning $c_2^{\rm DT} \in [-4.3 \times 10^{6}, 0]$.
    }
    \label{fig:lambda_tilde_DT}
\end{figure*}

While the individual tidal parameters $\Lambda_{2,A}$ yield only broad posterior distributions (see e.g. Fig.~\ref{fig:full_run} in Appendix~\ref{sec:app}), the binary tidal deformability $\tilde{\Lambda}$~\cite{Wade:2014vqa} is well constrained to $\tilde{\Lambda} = 294^{+14}_{-13}$ for the fiducial run at the 90\% credible interval, which is in excellent agreement with the injected value as shown in the left panel of Fig. \ref{fig:lambda_tilde_DT}. Similarly, we find that the quadrupolar dynamical phasing coefficient given by 
\begin{align}
\label{eq:DTcoeff}
c_2^{\rm DT} := -\frac{1}{X_1 X_2}&\Bigg[\frac{\Lambda_{2,1}}{\Omega_{2,1}^2} X_1^6(155-147X_1) \nonumber \\ 
&+ \frac{\Lambda_{2,2}}{\Omega_{2,2}^2}X_2^6 (155-147X_2)\Bigg],
\end{align}
where $X_A = m_A / M$, is found to be $c_2^{\rm DT} = -4.3^{+0.7}_{-0.7}\times 10^{5}$ at 90\% confidence, which is in excellent agreement with the injected value $c_2^{\rm DT} = -4.3 \times 10^{5}$. For comparison, the prior range is $c_2^{\rm DT} \in [-4.3 \times 10^{6}, 0]$ as shown in the right panel of Fig. \ref{fig:lambda_tilde_DT}. We find that $c_2^{\rm DT} = 0$ (adiabatic limit) is excluded at $>99\%$ confidence, showing a coherent measurement of dynamically driven dephasing within the signal.

\section{Conclusions}
\label{sec:Conclusions}
The next generation of ground-based GW detectors will have unprecedented sensitivities between $\sim 10$ Hz and a few kHz. The improved sensitivity at frequencies $\simeq 400$ Hz is particularly enticing as it will allow to place formidable constraints on the as-of-yet unknown EOS of neutron stars by measuring tidal effects in inspiralling neutron star binaries. Beyond the leading-order adiabatic tidal effects, dynamical tides can become important at higher frequencies and may be measureable in 3G detectors~\cite{Pratten:2019sed}. This would allow to independently verify EOS-independent relations~\cite{Chan:2014kua, Yagi:2016qmr, Godzieba:2020bbz} and search for higher-order phase transitions between the inspiral and post-merger regime~\cite{Weih:2019xvw} and therefore provide a means to study potentially new physics.

In this paper, we have investigated the detectability of dynamical tides from inspiralling BNS in 3G detectors. First, using the conservative distinguishability criterion~\cite{Lindblom:2008cm} we determined the SNRs required to disentangle dynamical from leading order adiabatic tides. We found that for loud enough signals, dynamical tides are indeed distinguishable but that the required SNR depends on the NS EOS, component masses and detector: binaries with a stiff EOS and low total mass present the best candidates for measuring dynamical tides. For a GW170817-like signal we found that irrespective of the EOS and the specific detector network, the distinguishability criterion is always fulfilled since the optimal SNR would be SNR 1031 (2935) for ET (CE) and the highest required SNR to distinguish dynamical tides is 84 and 102 respectively. 

We then applied this methodology to a population of BNS: We simulated 10,000 BNS inspiral signals with masses drawn from a double-peaked mass distribution consistent with the galactic double neutron star \emph{and} recycled pulsar distribution and three different EOS up to a redshift of $z=0.5$. 
We found that for the vast majority of signals no tidal information is recovered, and therefore they cannot be discerned from binary black holes unless an electromagnetic counterpart is detected simultaneously~\cite{Hinderer:2018pei}. 
In the case of the soft APR4 EOS tidal information can be obtained in only 4.0\% (10.2\%) of cases, for 3.7\% (4.2\%) dynamical tides can be measured in addition to adiabatic effects. 
However, given the BNS detection rates expected for 3G instruments of $10^3 - 10^5$ per year, for a significant number of BNS detections dynamical tides will be distinguishable. 

To understand whether simple distinguishability maps into a measurement of the $f$-mode frequencies of the two stars during the inspiral, we performed full Bayesian inference on a GW170817-like signal in a CE-ET detector network. While we found that the $f$-mode frequencies of the individual stars are only constrained to within a few hundred Hz, the leading-order PN phasing coefficient for dynamical tides $c_2^{\rm DT}$ is measured to $\mathcal{O}(30\%)$ and purely adiabatic tides (i.e. $c_2^{\rm DT} =0$) is excluded at $>99\%$ confidence. Fisher estimates promise tighter bounds on the $f$-mode frequencies, but we observe large correlations between the star's tidal deformability $\Lambda_A$ and its $f$-mode frequency, which are neglected in Fisher estimates. Similarly, the individual $\Lambda_A$ are poorly constrained, but the binary tidal deformability parameter $\tilde{\Lambda}$ is measured to within $\sim 5 \%$ at the 90\% CI. Preliminary studies suggest that similar to $\tilde{\Lambda}$, $c_2^{\rm DT}$ might be a preferred sampling parameter and we leave further investigations to future work. We leave a more comprehensive investigation across the BNS parameter space to future work.

The results presented here assume perfect knowledge of the waveform describing a BNS inspiral in General Relativity as well as the correctness of quasi-universal relations for tidal parameters. In addition, we assume perfect calibration knowledge when performing parameter estimation. All three assumptions are simplifications and hence pose a caveat to our analyses. Recent work suggests that systematic waveform errors will have the largest impact on astrophysical inference of the nuclear EOS while detector calibration errors still play a crucial but subdominant role~\cite{Chatziioannou:2021tdi, Pratten:2021pro, Essick:2022vzl}.
A future avenue we will explore is understanding how dynamical tidal information can be incorporated into the joint inference of the astrophysical population and the equation of state \cite{Golomb:2021tll}, especially in the presence of correlated parameters \cite{Biscoveanu:2021eht} and waveform systematics \cite{Pratten:2021pro}.
Another caveat to our analysis is the neglect of spin. The inclusion of spin will in practice affect the measurability of resonant tidal effects \cite{Ho:1998hq,Lai:2006pr,Flanagan:2006sb,Poisson:2016wtv,Ma:2020rak,Poisson:2020eki,Steinhoff:2021dsn,Kuan:2022etu}. For spins which are (anti-)aligned with the orbital angular momentum, the $f$-mode frequency is shifted upwards (downwards)~\cite{Doneva:2013zqa} and thus results in less (more) dephasing in the signal~\cite{Steinhoff:2021dsn} but we caution that rather high NS spins are needed to have a noticeable effect on the tidal phase~\cite{Ma:2020rak,Kuan:2022etu}. In addition, the inclusion of spin precession in the point-particle sector is known to break mass -- spin degeneracies, which results in a more accurate mass measurement~\cite{Vecchio:2003tn,Lang:2006bsg, Chatziioannou:2014coa,Pratten:2020igi} and therefore may improve the measurement of tidal parameters. We leave the extension to spinning neutron stars for future work.

\section*{Acknowledgments}
We thank Jocelyn Read for useful discussions and Richard O'Shaughnessy for comments on the manuscript as well as Sam Higginbotham for his contribution to the very early stages of this project.
N. W. and G. P. are supported by STFC, the School of Physics and Astronomy at the University of Birmingham and the Birmingham Institute for Gravitational Wave Astronomy. P. S. acknowledges support from STFC grant No. ST/V005677/1. Computations were performed using the University of Birmingham's BlueBEAR HPC service, which provides a High Performance Computing service to the University's research community, as well as resources provided by Supercomputing Wales, funded by STFC grants No. ST/I006285/1 and No. ST/V001167/1  supporting the UK Involvement in the Operation of Advanced LIGO.
Part of this research was performed while G.P. and P.S. were visiting the Institute for Pure and Applied Mathematics (IPAM), which is supported by the National Science Foundation (Grant No. DMS-1925919).
This manuscript has the LIGO document number P2200030.

\appendix

\section{Complete Parameter Estimation Results}
\label{sec:app}
\begin{figure*}
    \centering
    \includegraphics[width=\textwidth]{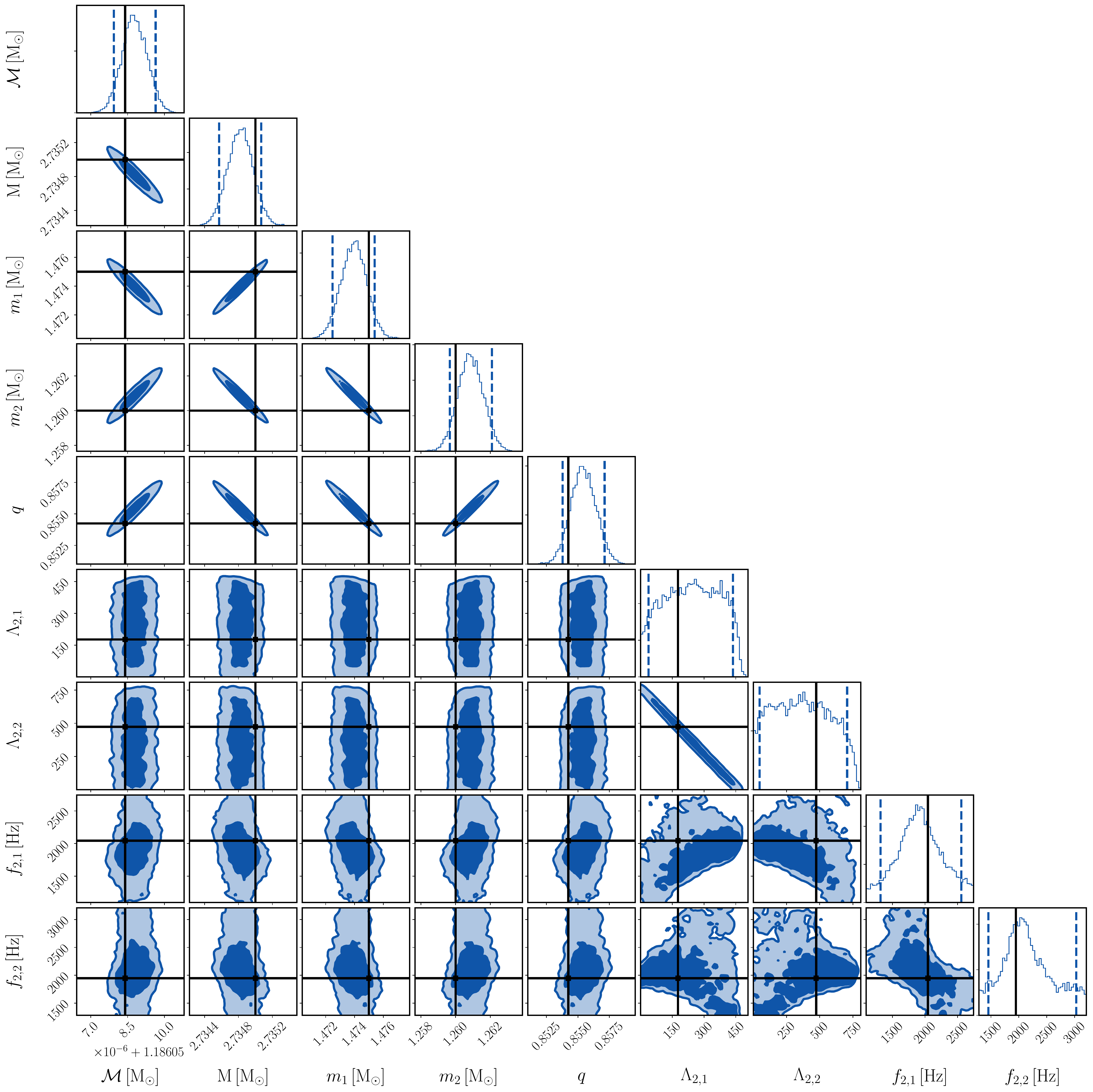}
    \caption{Corner plot showing the posterior distributions for all parameters of the fiducial analysis. The injected values (black lines) are shown alongside 50\% and 90\% contours for 2D posteriors, and 90\% confidence interval (blue dashed lines) for 1D posteriors.}
    \label{fig:full_run}
\end{figure*}

\begin{figure*}
    \centering
    \includegraphics[width=\textwidth]{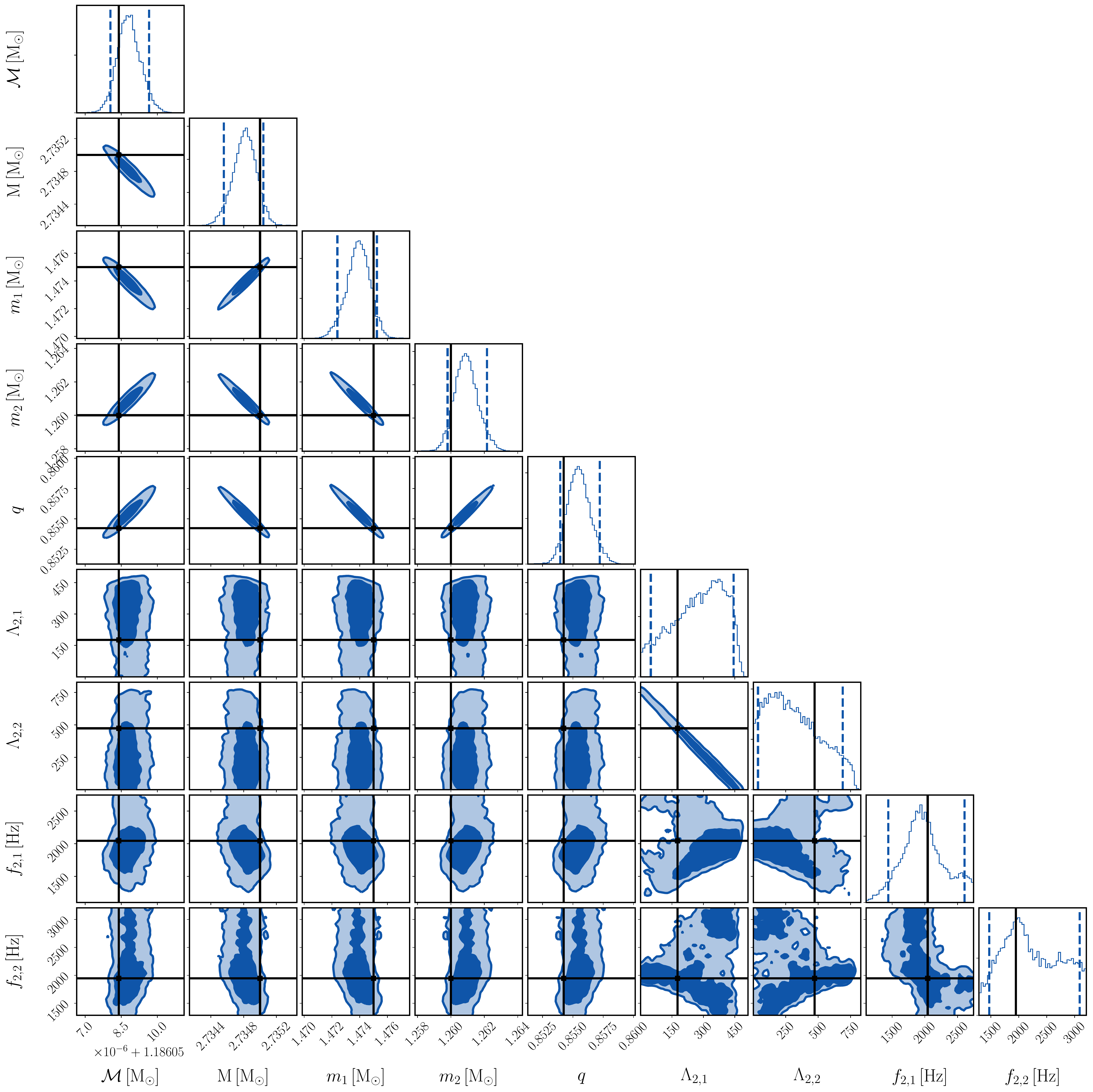}
    \caption{Corner plot showing the posterior distributions for all parameters of the analysis with the Gaussian mass prior. The injected values (black lines) are shown alongside 50\% and 90\% contours for 2D posteriors, and 90\% confidence interval (blue dashed lines) for 1D posteriors.}
    \label{fig:gaussian_run}
\end{figure*}

\begin{figure*}
    \centering
    \includegraphics[width=\textwidth]{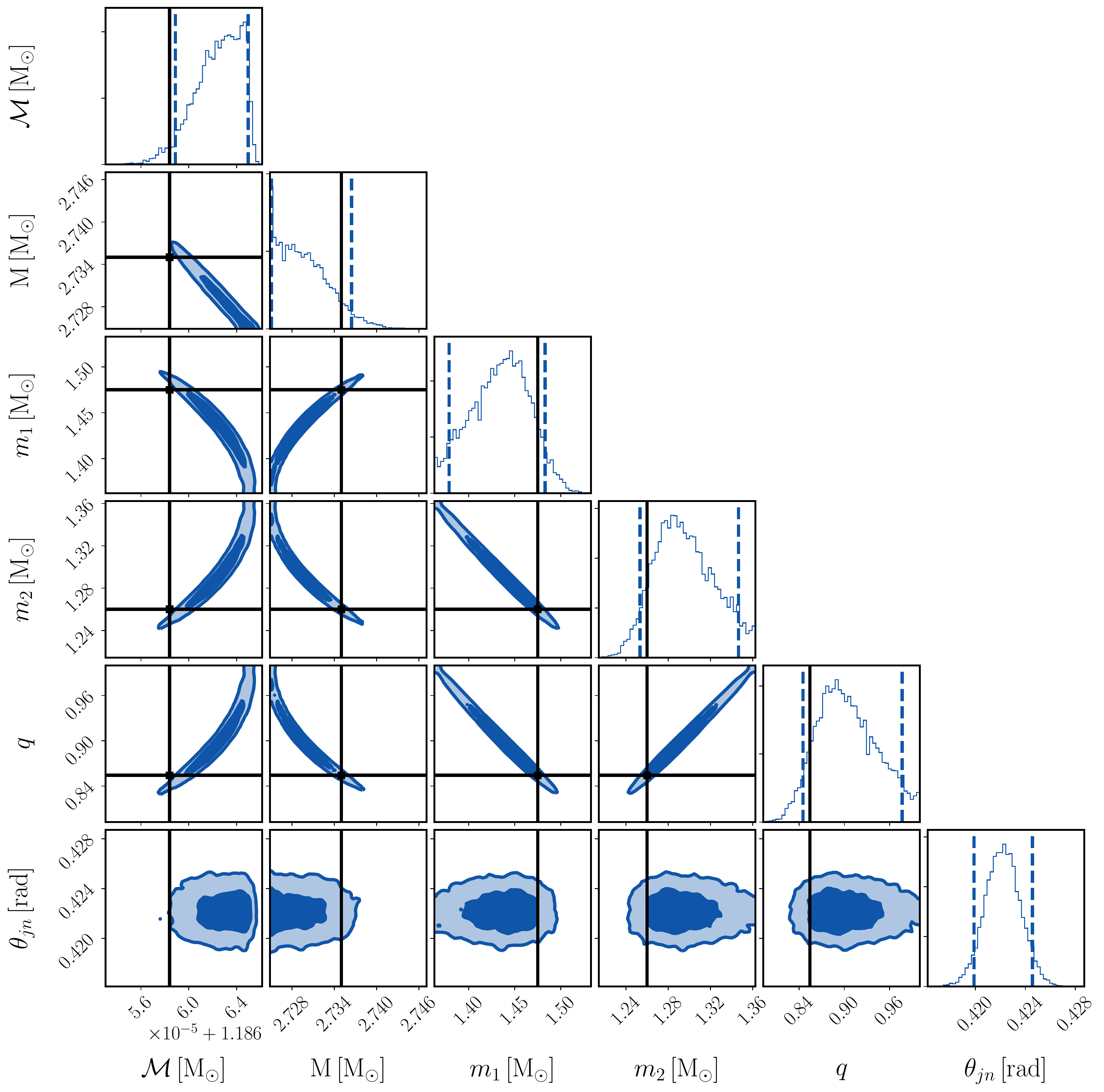}
    \caption{Corner plot showing the posterior distributions for all parameters from the 10-20 Hz non-tidal run which informed the Gaussian prior run. The injected values (black lines) are shown alongside 50\% and 90\% contours for 2D posteriors, and 90\% confidence interval (blue dashed lines) for 1D posteriors. $\theta_{jn}$ was not constrained correctly in this run, and the true value is not contained within the posterior.}
    \label{fig:10_20}
\end{figure*}

 Here we show the full 1- and 2D posterior distributions for all parameter estimation analyses presented in Sec.~\ref{sec:pe}. Due to fixing most of the extrinsic parameters, we show all intrinsic parameters as well as the angle between the total angular momentum and the line-of-sight, $\theta_{jn}$.The dashed vertical lines in the marginalised 1D distributions indicate the 90\% credible interval. All mass parameters are shown in the source frame. 
 The posteriors for the fiducial run are shown in Fig. \ref{fig:full_run}, whilst posteriors for the results containing Gaussian mass priors are shown in Fig. \ref{fig:gaussian_run} and the corresponding 10-20 Hz results to inform the Gaussian mass priors are shown in Fig. \ref{fig:10_20}. The injection parameters for all runs are as follows: $\mathcal{M} = 1.186M_{\odot}$, $M = 2.735 M_{\odot}$, $ m_1 = 1.475 M_{\odot}$, $ m_2 = 1.26 M_{\odot}$, $q = 0.854$, $\Lambda_{2,1} = 176.7$, $\Lambda_{2, 2} = 473.2$, $f_{2,1} = 2044 \, \rm Hz$, $f_{2,2} = 1947 \, \rm Hz$, $\theta_{jn} = 0.1 \, \rm rad$.

\bibliography{bibliography}

\end{document}